\documentclass[superscriptaddress,amsmath,amssymb,aps,prx,reprint,nofootinbib]{revtex4-2}

\usepackage[T1]{fontenc}
\usepackage[utf8]{inputenc}
\usepackage{braket}
\usepackage{siunitx}
\usepackage{graphicx}
\usepackage{dcolumn}
\usepackage{bm}
\usepackage{hyperref}
\usepackage{notes2bib}
\usepackage{siunitx}
\usepackage{bbold}
\usepackage[usenames,dvipsnames]{xcolor}

\definecolor{imec}{rgb}{0.02, 0.48, 0.32}
\definecolor{changed}{rgb}{0, 0, 0}
\definecolor{changed2}{rgb}{0, 0, 0}
\definecolor{todo}{rgb}{1, 0.75, 0.0}

\begin{document}

\title{Long coherence silicon spin qubit fabricated in a 300 mm industrial foundry}

\author{Petar Tomić}
\email{ptomic@phys.ethz.ch}
\affiliation{Laboratory for Solid State Physics, ETH Z\"{u}rich, CH-8093 Z\"{u}rich, Switzerland}
\author{Patrick B\"{u}tler}
\author{Yuze Wu}
\affiliation{Laboratory for Solid State Physics, ETH Z\"{u}rich, CH-8093 Z\"{u}rich, Switzerland}
\author{Bart Raes}
\author{Clement Godfrin}
\author{Stefan Kubicek}
\author{Julien Jussot}
\author{Yann Canvel}
\author{Yannick Hermans}
\author{Yosuke Shimura}
\author{Roger Loo}
\author{Sofie Beyne}
\author{Gulzat Jaliel}
\affiliation{IMEC, 3001 Leuven, Belgium}
\author{Thomas Van Caekenberghe}
\affiliation{IMEC, 3001 Leuven, Belgium}
\affiliation{Department of Electrical Engineering (ESAT), KU Leuven, Leuven, Belgium}
\author{Vukan Levajac}
\author{Danny Wan}
\affiliation{IMEC, 3001 Leuven, Belgium}
\author{Kristiaan De Greve}
\affiliation{IMEC, 3001 Leuven, Belgium}
\affiliation{Proximus Chair in Quantum Science and Technology, Department of Electrical Engineering (ESAT-MNS) KU Leuven, B-3001 Leuven, Belgium}
\author{Wister Wei Huang}
\affiliation{Laboratory for Solid State Physics, ETH Z\"{u}rich, CH-8093 Z\"{u}rich, Switzerland}
\affiliation{National University of Singapore, 119260 Singapore}
\author{Klaus Ensslin}
\author{Thomas Ihn}
\email{ihn@phys.ethz.ch}
\affiliation{Laboratory for Solid State Physics, ETH Z\"{u}rich, CH-8093 Z\"{u}rich, Switzerland}

\begin{abstract}
Silicon spin qubits offer long coherence times, a compact footprint and compatibility with industrial CMOS manufacturing. Here, we investigate spin qubits hosted in quantum dots fabricated in a state-of-the-art \SI{300}{\milli\meter} nanoelectronics foundry and demonstrate substantially enhanced coherence, achieving a Hahn-echo time of $T_2^{\text{Hahn}} = \SI{4}{\milli\second}$ for singlet–triplet oscillations. Employing noise spectroscopy and noise correlation measurements, we identify detuning noise with an amplitude of $\delta \varepsilon_{\text{rms}} = \SI{2.2}{\micro\electronvolt}$ (integrated over \SI{90}{\second}) and observe strong zero-phase correlations between two spatially separated spin qubits. The singlet--triplet basis intrinsically rejects these common-mode fluctuations, yielding a pronounced suppression of dephasing. Our results suggest that exploiting the versatility of silicon quantum dots to adapt the qubit encoding to the microscopic noise landscape represents a promising strategy for advancing scalable quantum information processing.
\end{abstract}

\maketitle

\section{Introduction}

The development of a large-scale quantum computer represents a significant challenge for both science and engineering. Among the various approaches to implementing qubits, spin qubits~\cite{loss1998quantum} confined in silicon (Si) metal–oxide–semiconductor (MOS) quantum dots have emerged as one of the promising technologies for achieving the scalability required for fault-tolerant quantum computing~\cite{kane1998silicon, zwanenburg2013silicon}. Their potential for large-scale integration arises from their compatibility with complementary MOS (CMOS) technology, which forms the foundation of the global semiconductor industry~\cite{vandersypen2019quantum, gonzalez2021scaling, maurand2016cmos, zwerver2022qubits}. This compatibility offers Si MOS spin-qubit architectures a realistic path toward integrating millions of qubits on a single chip, leveraging the massive investment in advanced fabrication techniques, materials, and infrastructure developed for classical computation.

Substantial progress has been made in demonstrating the essential building blocks for quantum computation in silicon, including high-fidelity single- and two-qubit gates~\cite{veldhorst2014addressable, yoneda2018quantum, yang2019silicon, wu2025simultaneous, xue2022quantum, noiri2022fast, mkadzik2022precision, mills2022two}, cryogenic CMOS based control~\cite{xue2021cmos, bartee2025spin}, long-range coupling via superconducting microwave resonators~\cite{harvey2022coherent}, and, more recently, high-fidelity spin shuttling~\cite{struck2024spin, xue2024si, de2025high}. Although the feasibility of fabricating spin qubits in industrial foundries has been demonstrated~\cite{maurand2016cmos, zwerver2022qubits, steinacker2025industry}, it remains unclear how far coherence times can be extended using state-of-the-art nanoelectronics fabrication facilities. Achieving long coherence times in foundry-fabricated silicon qubits is a crucial step toward realizing scalable quantum computing architectures~\cite{petit2020universal}.

\begin{figure*}[!t]
    \centering
    \includegraphics{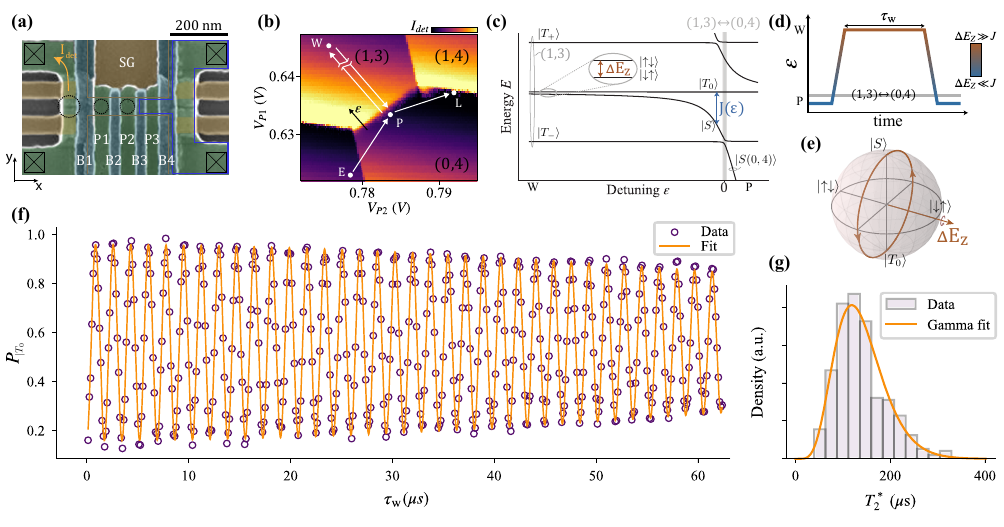}
    \caption{Device layout, charge stability, energy diagrams and states relevant for the singlet--triplet qubit, and characterization of singlet--triplet oscillations driven by Zeeman energy differences. (a) False-color scanning electron micrograph of the device. Gold, green, and blue indicate the first, second, and third gate layers, respectively. The black crossed squares are ohmic contacts. The left single-electron transistor (SET) functions as a charge detector, while the right one serves as a reservoir for the double quantum dot (region enclosed in dark blue). The double dot is defined beneath the plunger gates P1 and P2 (black dashed circles) within the channel shaped by the screening gates (SG, outlined in dark gold). (b) Detector current $I_{\textnormal{det}}$ as a function of plunger gate voltages $V_{P1}$ and $V_{P2}$, showing the charge stability diagram of a double dot in the (1,3)-(0,4) charge regime. The black arrow marks the detuning $\varepsilon$ axis, and the white arrows indicate the pulsing sequence used to generate $\ket{S}-\ket{T_0}$ oscillations. (c) Energy diagram of the singlet--triplet qubit relevant (1,3) and (0,4) charge states as a function of detuning at finite magnetic field. The avoided crossing arises from tunnel coupling between $\ket{S(1,3)}$ and $\ket{S(0,4)}$ states. The exchange energy $J(\varepsilon)$, corresponding to the energy difference between the hybridized singlet $\ket{S}$ and triplet $\ket{T_0}$ branches, is indicated in blue. In the far-detuned regime (W point), the states converge to $\ket{\downarrow \uparrow}$ and $\ket{\uparrow \downarrow}$ respectively, split by the Zeeman energy difference $\Delta E_\textnormal{Z}$, labeled in brown. (d) Detuning pulse sequence used to observe singlet–triplet oscillations. (e) Evolution of the quantum state at detuning point W on the Bloch sphere during the pulsing procedure in (d). (f) Probability of measuring $\ket{T_0}$ state as a function of time showing long lasting coherent $\ket{S}-\ket{T_0}$ oscillations.(g) Distribution of extracted $T_2^*$ times from (f), fitted with a Gamma distribution.
    }
    \label{fig:fig1}
\end{figure*}

In this work, we report the longest Hahn-echo ($T_2^\textnormal{Hahn}$) coherence times yet measured in gate defined quantum dot spin qubits fabricated entirely within an advanced \SI{300}{\milli\meter} CMOS foundry~\cite{stanoloss2025}. By driving coherent rotations along both axes of the singlet–triplet (S--T$_0$) qubit~\cite{levy2002universal, petta2005coherent} we extract key device parameters. We resolve distinct g-factor differences for ground and excited valley configurations, allowing us to extract the valley splitting and quantify the anisotropy of the spin-orbit interaction and its susceptibility to the valley degree of freedom. Furthermore, we characterize spin coherence, and quantify electrical noise in terms of detuning fluctuations~\cite{cywinski2008enhance, dial2013charge}. These measurements reveal a remarkably quiet electrical environment, with detuning noise amplitudes as low as \SI{2.2}{\micro\electronvolt} (integrated over \SI{90}{\second} per trace). Independent noise spectra from a neighboring single-electron transistor are consistent with this picture, yielding a power spectral density at \SI{1}{\hertz} of $\sqrt{S_0}\approx\SI{0.3}{\micro\electronvolt\per\sqrt{\hertz}}$. To understand the origin of the long coherence, we perform comprehensive noise spectroscopy measurements and cross-correlation analysis between neighboring spin qubits forming the S--T$_0$ qubits revealing a strong zero-phase correlation in the charge noise. This spatially correlated noise is generally considered a severe limitation in quantum computing, as it undermines the ability of standard error correcting codes~\cite{fowler2012surface, bombin2006topological} to suppress logical errors by scaling the code distance~\cite{mcewen2022resolving,google2025quantum,miao2023overcoming}. Remarkably, however, this in-phase correlated noise acts favorably in our system causing both spins to experience nearly identical fluctuations, making the S–T$_0$ encoding largely insensitive to such variations and strongly suppressing dephasing. These findings highlight that the impact of noise depends significantly on the choice of the qubit encoding. This insight suggests that exploiting the versatility of silicon quantum dots to adapt the qubit encoding to the microscopic noise landscape represents a promising strategy for advancing scalable quantum information processing.

\section{Device and qubit operation}

Fig.~\ref{fig:fig1}(a) presents the electron MOS quantum dot array we investigated in this work, fabricated on isotopically purified (400 ppm) undoped silicon. Three overlapping polysilicon gate layers define the confinement potential for the quantum dots and for the two charge detectors. The first gate layer serves as a screening gate, providing  confinement in the in-plane (y) direction. The second and third gate layers form the plunger (P1-P3) and barrier (B1-B4) gates, respectively. All gate layers are isolated via a \SI{5}{\nano\meter} SiO$_2$ layer while the Si/SiO$_2$ interface to the silicon substrate is formed by a \SI{20}{\nano\meter} thermally grown oxide. Furthermore, a shorted coplanar waveguide serving as an electron spin resonance (ESR) antenna (not shown in Fig.~\ref{fig:fig1}(a)) is positioned \SI{130}{\nano\meter} above the Si/SiO$_2$ interface.

\begin{figure*}[t]
    \centering
    \includegraphics{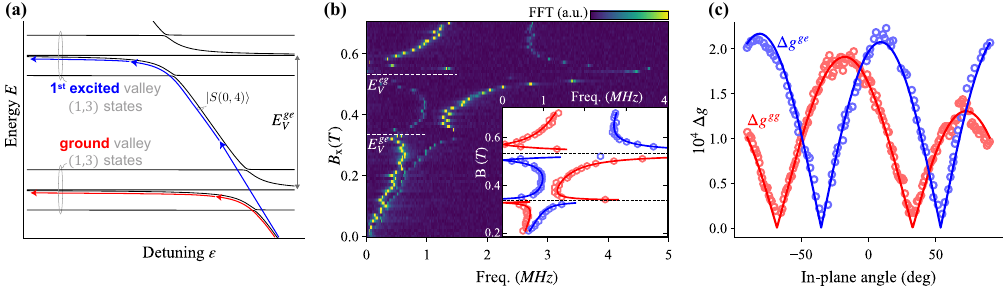}
    \caption{Excited-valley singlet--triplet qubit and $\Delta g$ anisotropy. Red (blue) denotes data or states associated with the ground (excited) valley throughout panels (a)–(c). (a) Energy levels as a function of detuning $\varepsilon$ at finite magnetic field, extending Fig.~\ref{fig:fig1}(c) to include both ground- and first-excited-valley (1,3) states. Arrows indicate initialization in the singlet states of ground $\ket{S^{gg}(1,3)}$ and excited $\ket{S^{ge}(1,3)}$ valley configuration. (b) Fourier transform of S--T$_0$ oscillations as a function of in-plane magnetic field $B_{\textnormal{x}}$ at fixed detuning position W. Two distinct frequency branches are visible corresponding to the ground- (dominant amplitude) and excited-valley (weaker amplitude) S--T$_0$ oscillations. White dashed lines mark anticrossings from spin-valley coupling at energies corresponding to valley-splitting $E_V^{ge}$ and $E_V^{eg}$. The inset shows model fits (solid lines) to the measured data (points). (c) $\Delta g$ values as a function of in-plane magnetic field angle $\phi$ (where $\phi=0$ corresponds to the $+x$ axis along $[110]$ in Fig.~\ref{fig:fig1}) extracted from S--T$_0$ oscillations of ground and excited valley states. Circles are data, and solid lines are model fits.
    }
    \label{fig:fig2}
\end{figure*} 

Two quantum dots are formed below gates P1 and P2 in Fig.~\ref{fig:fig1}(a). The rightmost side below gates P3, B4, and including the right charge detector, is operated in accumulation mode to serve as the electron reservoir (the area enclosed by the blue outline in Fig.~\ref{fig:fig1}(a)). The left charge detector, operated in DC mode with a bandwidth of \SI{30}{\kilo\hertz}, is used for implementing a single-shot spin readout. The double dot is tuned to the (1,3) - (0,4) charge configuration [charge stability diagram shown in Fig.~\ref{fig:fig1}(b)], while the barriers were tuned to enable latched spin readout~\cite{harvey2018high}. In this regime, tunnel coupling to the detector was significantly smaller than inter-dot coupling, which was smaller than coupling to the electron reservoir. The three plunger gates P1-P3 are connected via bias-tees to both DC sources and arbitrary waveform generator (AWG) outputs for high-frequency pulsing. All measurements were performed in a dilution refrigerator at a base temperature of \SI{8}{\milli\kelvin}.

We implement a singlet--triplet qubit in this device. This qubit is encoded in the two-electron spin subspace of a double quantum dot, spanned by the charge hybridized singlet $\ket{S} = \alpha \ket{S(1,3)} + \beta \ket{S(0,4)}$ and the triplet $\ket{T_0(1,3)}$ states. The energy difference between these two states defines the exchange energy $J$. By tuning with gate voltages the difference between the electrochemical potentials in each of the dots, called detuning $\varepsilon$, one can control the degree of hybridization between $\ket{S(1,3)}$ and $\ket{S(0,4)}$. The hybridization lowers the energy of $\ket{S}$ below that of $\ket{T_0(1,3)}$, and hence controls the exchange energy $J(\varepsilon)$, as illustrated in Fig.~\ref{fig:fig1}(c).

The second important energy scale, $\Delta E_\textnormal{Z} = \Delta g \mu_\textnormal{B} B$, arises from the difference in local Zeeman energies of the two dots in the presence of an external magnetic field $B$. This energy difference is governed by the $g$-factor difference $\Delta g$, which originates from variations in the spin-orbit interaction between the two dot locations. This gradient defines the coupling strength between $\ket{S}$ and $\ket{T_0}$. However, it is not tunable by detuning $\varepsilon$. Combined with exchange modulations it allows for full single-qubit control. The effective Hamiltonian in the $\{\ket{S}, \ket{T_0}\}$ basis is
\begin{equation}
    H = \frac{1}{2} J(\varepsilon) \sigma_z + \frac{1}{2} \Delta E_\textnormal{Z}(B) \sigma_x.
    \label{st_hamiltonian}
\end{equation}
For noise spectroscopy measurements, single-spin rotations were driven using the on-chip ESR antenna [see Appendix~\ref{app:noise}].

The gate pulsing sequence is outlined in the charge-stability diagram shown in Fig.~\ref{fig:fig1}(b). Initialization is achieved by loading the $(0,4)$ charge configuration ground state $\ket{S(0,4)}$ via a rate-limited voltage ramp from the $(0,3)$ to the $(0,4)$ charge state to ensure ground-state occupation (E $\rightarrow$ P). This is followed by detuning sweeps (P $\rightarrow$ W) through the $(1,3)-(0,4)$ transition to prepare either $\ket{S(1,3)}$ via rapid adiabatic passage or $\ket{\uparrow \downarrow (1,3)}$ via slow adiabatic passage~\cite{taylor2007relaxation}. The state is read out via the inverse sweep (W $\rightarrow$ P) followed by a pulse (P $\rightarrow$ L) to a latching configuration.

\section{Characterizing the S--T$_0$ qubit}
\subsection{Determination of $\Delta E_\textnormal{Z}$}

The coherent dynamics of the S--T$_0$ qubit are driven using the detuning pulse shown in Fig.~\ref{fig:fig1}(d). The qubit is prepared at detuning point W in the $\ket{S(1,3)}$ state by rapid adiabatic passage from detuning point P deep into the $(1,3)$ charge configuration. After preparation, the detuning is held at W for a waiting time $\tau_\textnormal{w}$, during which the qubit evolves under a Hamiltonian dominated by the Zeeman energy difference ($\Delta E_Z \gg J$) inducing coherent oscillations between $\ket{S}$ and $\ket{T_0}$ states. The evolution of the qubit state during $\tau_\textnormal{w}$ is shown on the Bloch sphere in Fig.~\ref{fig:fig1}(e). At the end of this interval, the spin state is projected onto a charge configuration for readout by returning to point P. In Fig.~\ref{fig:fig1}(f) we plot the probability of observing the $\ket{T_0}$ state $P_{T_0}$ after the pulse as a function of waiting time and observe long lasting coherent oscillations with a frequency of $\SI{0.58}{MHz}$ between $\ket{S}$ and $\ket{T_0}$ states at an in-plane magnetic field of $B_x=\SI{200}{\milli\tesla}$.

\begin{figure*}[t]
    \centering
    \includegraphics{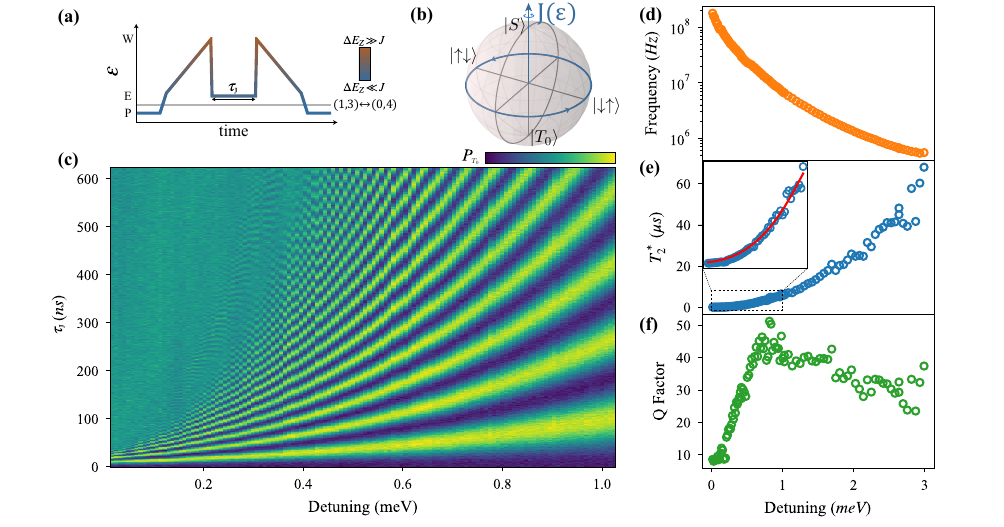}
    \caption{Characterizing exchange interaction. (a) Detuning pulse sequence used to drive exchange oscillations. (b) Evolution of the quantum state on the Bloch sphere during exchange time $\tau_J$ following the pulsing procedure in (a). (c) Probability of measuring $\ket{T_0}$ state as a function of exchange time $\tau_J$ and detuning $\varepsilon$, showing coherent exchange oscillations. (d-f) Extracted frequency, dephasing time $T_2^*$, and quality factor Q as a function of detuning, obtained from fits to the oscillations in (c). The inset in (e) shows fit (red line) to a dephasing model based on detuning noise.}
    \label{fig:fig3}
\end{figure*}

To accurately quantify coherence, we acquire multiple oscillation traces, each with a total integration time of \SI{87}{\second}. We observe that the extracted dephasing time $T_2^*$ is not constant but fluctuates significantly between consecutive measurements, as illustrated by the histogram in Fig.~\ref{fig:fig1}(g). This behavior indicates that the dominant noise source evolves on time scales longer than the single-trace acquisition time, placing the experiment in the non-ergodic regime. In this limit, a single measurement is not representative of the ensemble average and $T_2^*$ becomes a random variable with a distribution depending on the total measurement time~\cite{nonergodic_noise}. The resulting distribution of $T_2^*$ values is well-fitted by a Gamma distribution, with a maximum at $T_2^* = \SI{118}{\micro\second}$.
% Interestingly, even after \SI{8}{\hour} of measurements we observed strong beating and clear oscillations for evolution times of more then \SI{100}{\micro\second}.

To tune the qubit to an operating point where the logical subspace is isolated from spin-valley mixing, we map the energy spectrum by measuring S--T$_0$ oscillations as a function of in-plane magnetic field. The Fourier transform of these oscillations as a function of $B_x$, shown in Fig.~\ref{fig:fig2}(b), reveals not one, but two distinct frequency components. The dominant frequency component corresponds to the S--T$_0$ oscillations where both dots are in their respective valley ground state. We label this configuration with $gg$ -- where the notation denotes the valley state ($g$round or $e$xcited) of the left and right dot, respectively. The preparation of this state is represented by the red arrows in Fig.~\ref{fig:fig2}(a). The second, weaker frequency component are S--T$_0$ oscillations as well. They correspond to a valley configuration where either of the dots is in the first excited valley configuration ($ge$ - we assume lower excitation energy in a dot with three electrons). This excited valley S--T$_0$ qubit is prepared with a small probability (hence the weaker signal in the spectrum) during the state preparation sweep (P $\rightarrow$ W in Fig.~\ref{fig:fig1}(b)) when non-adiabatic transitions occur, as illustrated by the blue arrows in Fig.~\ref{fig:fig2}(a). The two frequencies in the spectrum stem from the different orbital characters of the ground and excited valley configurations, which alter the spin-orbit strength and hence lead to distinct g-factor differences~\cite{veldhorst2015spin, ruskov2018electron} $\Delta g^{gg}$ and $\Delta g^{ge}$.

As the magnetic field increases, we observe in Fig.~\ref{fig:fig2}(b) avoided crossings in the spectrum. These arise from spin-valley coupling, which hybridizes the S--T$_0$ states with the $\ket{T_-}$ and $\ket{T_+}$ states of different valley configurations~\cite{jock2022silicon}. In the inset of Fig.~\ref{fig:fig2}(b) we fit the peak frequencies of the spectrum to a model incorporating spin-valley interactions (for details see Appendix~\ref{app:spin_valley}). From the fit, we extract the valley splitting $E_V^{ge}=\SI{39}{\micro\electronvolt}$ and $E_V^{eg}=\SI{62}{\micro\electronvolt}$, the g-factor differences $\Delta g^{gg}=\SI{1.6e-4}{}$, $\Delta g^{ge}=\SI{2.5e-4}{}$, and the spin-valley couplings $\Delta^{(1)}=\SI{15}{\mega\hertz}$ and $\Delta^{(2)}=\SI{37}{\mega\hertz}$. These relatively small valley-splittings are consistent with the device's thick $\SI{20}{\nano\meter}$ oxide. For a fixed electron occupation, this geometry results in a reduced vertical electric field compared to thin-oxide devices, leading to weaker confinement normal to the interface~\cite{friesen2007valley}.

Having identified a magnetic field magnitude range where the logical subspace is isolated from valley mixing, we next explore the dependence of $\Delta E_\textnormal{Z}$ on the in-plane field angle $\phi$. Since the spin-orbit interaction induces an anisotropy in $\Delta g$, optimizing the field orientation is essential for the addressability of individual spins---a requirement for the noise spectroscopy performed in Sec.~\ref{sec:noise_characterization}. Furthermore, mapping this anisotropy provides detailed insight into the spin-orbit physics of the device. We therefore measure both $\Delta g^{gg}$ and $\Delta g^{ge}$ by sweeping the in-plane angle of the magnetic field at the fixed absolute value $|\vec{B}| = \SI{250}{\milli\tesla}$ [see Fig.~\ref{fig:fig2}(c)]. We fit this data to a model that accounts for both Rashba and Dresselhaus spin-orbit interactions~\cite{jock2018silicon}
\begin{equation}
    \Delta g \left( \phi \right) = \frac{2}{\mu_B} \left| \Delta \alpha - \Delta \beta \sin\left[2 \left( \phi - \phi_\textnormal{offset}\right)\right]  \right| 
\end{equation}
where $\Delta \alpha$ and $\Delta \beta$ are differences in the Rashba and Dresselhaus spin-orbit strength between the dots respectively, and $\phi$ is the in-plane angle of the magnetic field where $\phi=0$ corresponds to the $+x$ axis (along $[110]$) in Fig.~\ref{fig:fig1}(a). For $\Delta g^{gg}$ we extract $\Delta \alpha = \SI{210}{\kilo\hertz\per\tesla}$ and $\Delta \beta = \SI{1.13}{\mega\hertz\per\tesla}$ and for $\Delta g^{ge}$ we extract $\Delta \alpha = \SI{33}{\kilo\hertz\per\tesla}$ and $\Delta \beta = \SI{1.47}{\mega\hertz\per\tesla}$. With these measurements, we uniquely isolate the influence of the valley degree of freedom on spin-orbit interactions. The extracted parameters are about an order of magnitude smaller than in previous reports~\cite{jock2018silicon, tanttu2019controlling, cifuentes2024bounds, chittock2025radio}. This finding is consistent with the weaker vertical confinement~\cite{jock2018silicon} stemming from the thick oxide, which also yielded the small valley splitting. However, the physical origin of the observed phase difference in $\Delta g$ between the two valley configurations remains unclear and cannot be accounted for within our current understanding of the interplay between interface roughness, spin–orbit coupling~\cite{cifuentes2024bounds}, and valley physics, indicating that further investigation is needed.

\subsection{Calibrating exchange}
\label{sec:exchange}

We probe the second rotation axis of the qubit by inducing exchange oscillations using the detuning pulse sequence shown in Fig.~\ref{fig:fig3}(a). We use slow adiabatic passage [P $\rightarrow$ W, see Fig.~\ref{fig:fig1}(b)] to prepare the $\ket{\uparrow \downarrow}$ state in point W deep in the (1,3) charge configuration, followed by a diabatic sweep to a fixed detuning point E where $\Delta E_\textnormal{Z} \ll J$ to induce exchange oscillations between $\ket{\uparrow \downarrow}$ and $\ket{\downarrow \uparrow}$. The evolution of the qubit state during exchange time $\tau_\textnormal{J}$ is shown on the Bloch sphere in Fig.~\ref{fig:fig3}(b). The resulting exchange oscillations as a function of detuning and exchange time are plotted in Fig.~\ref{fig:fig3}(c). By adjusting the detuning $\varepsilon$ at point E, the oscillation frequency can be continuously varied---from negligible values in the regime $J \ll \Delta E_\textnormal{Z}$ to as high as \SI{200}{\mega\hertz} [see Fig.~\ref{fig:fig2}(d)].

\begin{figure}[t]
    \centering
    \includegraphics{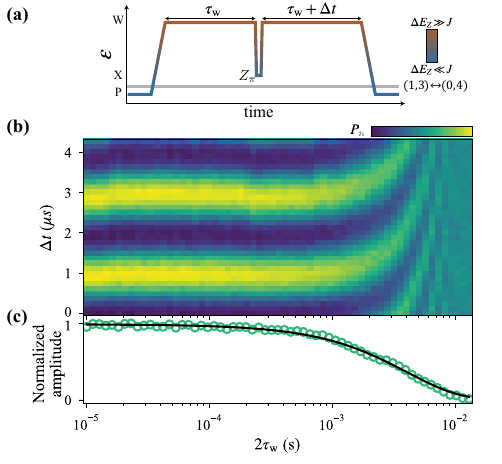}
    \caption{Hahn-echo experiment. (a) Detuning pulse sequence used for the Hahn-echo experiment. An exchange $\pi$-pulse is applied to refocus dephasing arising from fluctuations in $\Delta E_{\textnormal{Z}}$. (b) Probability of measuring the $\ket{T_0}$ state as a function of total waiting time $2 \tau_{\textnormal{W}}$ and additional delay $\Delta t$, measured using the sequence in (a). The phase shift of the oscillations at larger total waiting times originates from a combination of finite bias-tee response time and the $\Delta g$ susceptibility to plunger gate voltages. (c) Normalized oscillation amplitude extracted from (b) as a function of total wait time. The data (green circles) are fitted with an exponential decay $\exp (-2 \tau_{\textnormal{W}} / T_2^{\textnormal{Hahn}} )$, yielding $T_2^{\textnormal{Hahn}} = \SI{4}{\milli\second}$.}
    \label{fig:fig4}
\end{figure}

The decay of these oscillations is charge-noise dominated and can be viewed as the effect of detuning noise characterized by the root-mean-squared noise $\delta \varepsilon_\textnormal{rms}$~\cite{dial2013charge}. By modeling the dephasing time $T_2^*$ as a function of detuning [Fig.~\ref{fig:fig3}(e)] in the exchange-driven regime ($\Delta E_\textnormal{Z} \ll J$), we extract $\delta \varepsilon_\textnormal{rms}$ from~\cite{dial2013charge}
\begin{equation}
    T_2^* = \left( \frac{\pi\sqrt{2}}{h}  \frac{\mathrm{d}J(\varepsilon)}{\mathrm{d} \varepsilon} \delta\varepsilon_{\text{rms}} \right)^{-1}.
    \label{eq:t2_vs_det} 
\end{equation}
In the inset of Fig.~\ref{fig:fig3}(e), we fit our experimental data, obtained over a measurement time of approximately $\SI{90}{\second}$ per trace, with Eq.~\ref{eq:t2_vs_det} yielding a detuning noise of $\delta \varepsilon_\textnormal{rms} = \SI{2.2}{\micro\electronvolt}$. Although direct comparison with literature values is complicated by differences in measurement duration and underlying noise spectra, this result indicates a remarkably low level of charge noise~\cite{petersson2010quantum, jirovec2021singlet, thorgrimsson2017extending, chittock2025radio, jock2018silicon, kranz2020exploiting}. In combination with SET noise measurements yielding a power spectral density at \SI{1}{\hertz} of $\sqrt{S_0} \approx \SI{0.3}{\micro\electronvolt\per\sqrt{\hertz}}$ (see Appendix~\ref{app:set_noise}), these findings point to an exceptionally clean and stable electrostatic environment, highlighting the high quality of the industrial fabrication process~\cite{elsayed2024low}.

A more direct benchmark related to gate fidelities is the $Q$-factor, which indicates how many coherent oscillations fit within $T_2^*$. We achieve $Q$-factors peaking around 50 for an exchange frequency of \SI{10}{\mega\hertz} [see Fig.~\ref{fig:fig3}(f)]. Crucially, this result is obtained at finite detuning, where the qubit is first-order sensitive to charge noise. Transitioning to a symmetric operation point would suppress this sensitivity to second order~\cite{reed2016reduced, martins2016noise}, a strategy expected to yield a substantial enhancement in coherence. This indicates that the platform is capable of reaching state-of-the-art $Q$-factors on the order of hundreds~\cite{weinstein2023universal, ha2025two}.

\subsection{Spin-echo measurements}

Having established and calibrated the two rotation axes of the S--T$_0$ qubit, we suppress dephasing through Hahn-echo measurements and, by overcoming the long noise-autocorrelation time, restore the ergodic averaging regime. After having prepared the state $\ket{S(1,3)}$ at point W, we use the exchange $\pi$-pulse at point X, implemented with the detuning pulse sequence shown in Fig.~\ref{fig:fig4}(a), to reverse the phase accumulation caused by low-frequency noise in $\Delta E_\textnormal{Z}$. Figure~\ref{fig:fig4}(b) shows the probability of measuring the state $\ket{T_0}$ as a function of the total waiting time $2 \tau_\textnormal{w}$ at point W and an additional delay $\Delta t$ applied in the second half of the pulse, revealing coherent oscillations that persist well into the millisecond regime. At large $\tau_\textnormal{w}$, the oscillations also exhibit a phase shift due to the finite bias-tee time constant and the sensitivity of $\Delta E_\textnormal{Z}$ to plunger gate voltages (see Appendix~\ref{app:echo_phase}). The decay of oscillation amplitudes, extracted from Fig.~\ref{fig:fig4}(b), is plotted in Fig.~\ref{fig:fig4}(c) as a function of the total waiting time and fitted with the exponential envelope $\exp(-2 \tau_{\textnormal{w}} / T_2^{\text{Hahn}})$ yielding a coherence time of $T_2^\textnormal{Hahn} = \SI{4}{\milli\second}$. To the best of our knowledge, this is the largest $T_2^{\text{Hahn}}$ value reported in gate defined quantum dot spin qubits to date~\cite{stanoloss2025, steinacker2025industry}. The substantial improvement over $T_2^*$ confirms that decoherence is dominated by low-frequency noise and demonstrates the exceptional device quality achievable with industrial CMOS fabrication.

\section{Noise characterization}
\label{sec:noise_characterization}

\begin{figure}[!t]
    \centering
    \includegraphics{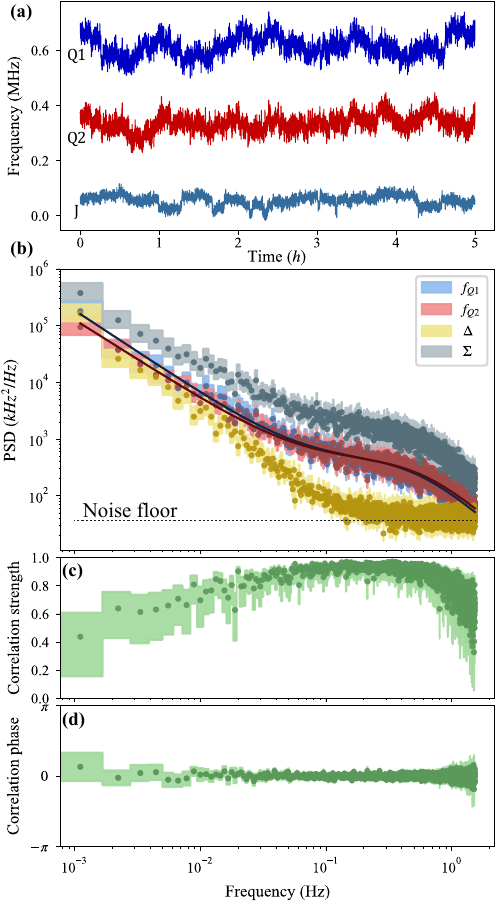}
    \caption{Noise characterization. (a) Time traces of frequencies of single-spin qubit $f_{Q1}$ and $f_{Q2}$, and the exchange frequency $f_\textnormal{J}$ (blue, red and light-blue trace, respectively), shifted by arbitrary offsets for clarity. (b) Power spectral densities (PSDs) of $f_{Q1}$ (blue) and $f_{Q2}$ (red) with fits (solid lines) to a model combining $1/f^{\alpha}$ noise from weak coupling to an ensemble of two-level-fluctuators (TLFs) and a Lorentzian component from a single strongly coupled TLF. PSDs of the sum $\Sigma=f_{Q1} + f_{Q2}$ (grey) and the difference $\Delta = f_{Q1} - f_{Q2}$ (gold) are also shown. Throughout the measured frequency range, $\Delta$ --- corresponding to Zeeman energy difference --- exhibits significantly lower noise level than the individual spins. (c) Correlation strength and (d) correlation phase between noise signals $f_{Q1}$ and $f_{Q2}$ indicating strong in-phase correlations over the entire frequency range. The correlation strength peaks at the frequencies where the single TLF dominates the individual spin qubit PSDs. In panels (b)–(d), the shaded regions represent 90\% confidence intervals.}
    \label{fig:fig5}
\end{figure}

The long coherence times observed for both rotation axes of the qubit motivate a detailed investigation of the noise performance of the device. We explore the noisy environment of the qubit by interleaving single-shot measurements to simultaneously track both single-spin qubit frequencies $f_\textnormal{Q1}$ and $f_\textnormal{Q2}$ (via ESR) and exchange frequency $f_\textnormal{J} \equiv J/h$ (as described in Sec.~\ref{sec:exchange}). We employ a Bayesian estimation procedure~\cite{yoneda2023noise} to enhance the effective sampling rate of these interleaved measurements to $f_\textnormal{Sampling} = \SI{3}{\hertz}$. Figure~\ref{fig:fig5}(a) shows extracted time traces of $f_\textnormal{Q1}$, $f_\textnormal{Q2}$ and $f_\textnormal{J}$ acquired continuously over a period of \SI{5}{\hour}. Further experimental details are provided in the Appendix~\ref{app:noise}.

We calculate the power spectral density (PSD) $S(f)$ of the two single-spin qubit frequencies $S_\textnormal{Q1}$, $S_\textnormal{Q2}$, and of the exchange frequency $S_J$ using the Bayesian estimation method described in Ref.~\cite{gutierrez2022bayesian}. The power spectral densities of both qubits (red and blue traces in Fig.~\ref{fig:fig5}(b)) exhibit a characteristic $1/f^{\alpha}$ dependence at low frequencies, with a deviation of this trend at higher frequencies. The PSD data can be described by
\[ S(f) = a \cdot 1/f^{\alpha}  + b  \cdot \gamma / (\gamma^2 + f^2), \]
where $\alpha=1.4$ and $\gamma=\SI{0.6}{\hertz}$. This amounts to
modeling the noise as originating from weak coupling to an ensemble of two-level fluctuators (TLFs) producing a $1/f^{\alpha}$ background, together with an uncorrelated single, strongly coupling TLF that contributes a Lorentzian component to the spectrum~\cite{elsayed2024low, yoneda2023noise, paladino20141}.
% Check in literature if my data is much lower in noise.

It is instructive to calculate the PSDs of the sum, $\Sigma = f_\textnormal{Q1} + f_\textnormal{Q2}$, and the difference, $\Delta = f_\textnormal{Q1} - f_\textnormal{Q2}$, of the two qubit frequencies. The difference $\Delta$ corresponds to the effective Zeeman splitting difference $\Delta E_\textnormal{Z}$ relevant for the S--T$_0$ qubit. The resulting PSDs, $S_{\Sigma}(f)$ and $S_{\Delta}(f)$, are shown in Fig.~\ref{fig:fig5}(b) (gray and gold traces, respectively). Compared to the individual qubit spectra, $S_{\Sigma}$ exhibits a higher noise level across the entire measured frequency range. In contrast, $S_{\Delta}(f)$ is suppressed by more than an order of magnitude relative to the individual-qubit PSDs in the regime where the measured spectrum hits the estimation noise floor, approaching the individual noise traces only at the lowest frequencies ($f\lesssim \SI{1e-2}{\hertz}$). The apparent flattening of $S_{\Delta}(f)$ at higher frequencies reflects the noise floor imposed by the Bayesian estimation procedure used to extract the time traces. The striking reduction of noise in $S_{\Delta}(f)$ demonstrates that the S--T$_0$ qubit experiences a significantly quieter noise environment than the individual spin qubits from which it is formed.

The reason for such a large difference between $S_{\Sigma}(f)$ and $S_{\Delta}(f)$ arises from correlated noise acting on the two spins. The PSDs can be expressed as
\begin{equation}
    S_{\Sigma/\Delta}(f) = S_\textnormal{Q1}(f) + S_\textnormal{Q2}(f) \pm 2 \operatorname{Re}\{S_\textnormal{Q1,Q2}(f)\}
    \label{psd_q1q2}
\end{equation}
where $S_\textnormal{Q1,Q2}(f) \in \mathbb{C}$ is the complex-valued cross-correlation PSD function of the noise acting on the two qubits. In the absence of correlations, $S_{\Sigma}(f)$ and $S_{\Delta}(f)$ would be the same and equal to the sum of the individual PSDs. The observed suppression of $S_{\Delta}(f)$ relative to $S_{\Sigma}(f)$ implies that $\operatorname{Re}\{S_\textnormal{Q1,Q2}(f)\} > 0$, and an in-phase correlation between the noise affecting the two qubits.

We quantify the degree of correlation by evaluating the complex linear correlation coefficient defined as $C_\textnormal{Q1,Q2}(f) = S_\textnormal{Q1,Q2}(f) / \sqrt{S_\textnormal{Q1}(f)S_\textnormal{Q2}}(f)$. We plot the magnitude and phase of $C_\textnormal{Q1,Q2}(f)$ as a function of frequency in Figs.~\ref{fig:fig5}(c) and (d) respectively. Across the entire measured frequency range, we observe a large correlation magnitude and a correlation phase close to zero. The high degree of correlation points to electrical noise as the dominant noise source, since hyperfine noise would not give such strong correlations over the approximately \SI{80}{\nano\meter} inter-dot separation~\cite{yoneda2023noise}.

Interestingly, the strongest correlations occur in the frequency range where the Lorentzian component dominates the individual PSDs, suggesting that a single TLF contributes simultaneously to the noise of both qubits. This correlated charge noise acts as a common-mode fluctuation. While individual spins are highly susceptible to this noise, the singlet–triplet qubit is sensitive only to the difference in the Zeeman energy. As a result, the strong in-phase correlation effectively decouples the qubit from the dominant noise source in the environment. Consequently, the long coherence times reported here result from the synergy of two factors: the pristine electrostatic environment achieved through industrial fabrication and the favorable in-phase noise correlations that further suppress dephasing in the S--T$_0$ basis.

Finally, we observe no correlation, within the uncertainty of our analysis, between individual qubit frequencies and exchange oscillations (see Appendix~\ref{app:exchange_correlations}). This lack of correlation likely arises from the strong electric field dependence of TLF frequencies~\cite{lisenfeld2023enhancing}. Since the ESR measurement for individual qubit frequencies and exchange measurements are performed at different detuning points, the relevant fluctuators may couple differently in each detuning regime, thereby suppressing any direct correlation.

\section{Conclusion}

In summary, we have demonstrated a long-lived singlet--triplet qubit in a gate defined silicon MOS double quantum dot fabricated in an advanced industrial \SI{300}{\milli\meter} CMOS foundry. Using this platform, we achieved a Hahn-echo coherence time of $T_2^{\text{Hahn}} = \SI{4}{\milli\second}$, the longest reported $T_2^{\text{Hahn}}$ time for a gate-defined quantum dot spin qubit. Such a long coherence is primarily attributed to the remarkably low noise electrical environment enabled by the state-of-the-art fabrication process. This is quantitatively confirmed by detuning noise $\delta \varepsilon_\textnormal{rms} = \SI{2.2}{\micro\electronvolt}$, measured in $\SI{90}{\second}$ per trace, in combination with nearby SET noise spectroscopy yielding a power spectral density at \SI{1}{\hertz} of $\sqrt{S_0} \approx \SI{0.3}{\micro\electronvolt\per\sqrt{\hertz}}$.

While this low-noise environment provides a crucial baseline, correlated noise spectroscopy revealed an additional factor responsible for the qubit's longevity. We observed strong in-phase correlations in charge noise affecting the spins within the two quantum dots. The singlet--triplet basis, being sensitive to the difference in the Zeeman energies, provides an intrinsic noise-filtering mechanism against common-mode fluctuations. The effectiveness of this filtering is directly proportional to the strength of the noise correlation.

These findings highlight the intrinsic flexibility of semiconductor spin qubits as a promising resource for enhancing coherence. Because silicon quantum dot arrays can support a broad spectrum of qubit encodings---spanning from single-spin Loss--DiVincenzo modes to composite multi-electron subspaces like singlet--triplet and exchange-only qubits~\cite{burkard2023semiconductor}---on essentially identical device layouts, the operational mode can be chosen to exploit the specific correlation properties of the local noise environment. Our results demonstrate this principle. Although in-phase charge fluctuations limit single-spin coherence, they form a robust decoherence-free subspace for the singlet--triplet encoding. Furthermore, the strong electric-field dependence of two-level fluctuators suggests that in-situ tuning of gate voltages could be used to navigate to electrostatic ''sweet spots'' where differential-mode noise is minimized. Future work should investigate the feasibility of such active noise-correlation tuning and probe the temporal stability of these correlations. Ultimately, the ability to adapt the qubit definition to the microscopic noise landscape identifies a compelling route for optimizing coherence in large-scale silicon quantum processors.

% \vspace{1em}

% The data supporting the findings of this work will be made available via the ETH Research Collection.

% \textcolor{changed2}{All data used in this Letter is made available online at the ETH Z\"{u}rich research collection \href {https://doi.org/10.3929/ethz-b-000508511}{DOI 10.3929/ethz-b-000508511}.} \textcolor{red}{Get new DOI}.

\begin{acknowledgments}
We acknowledge the support from Peter Maerki and Thomas Baehler. This work was supported as a part of NCCR SPIN, a National Centre of Competence in Research, funded by the Swiss National Science Foundation (grant number 225153). P.B. acknowledges support from the QuantERA project.

\end{acknowledgments}

\appendix
\section{Hahn echo phase shift}
\label{app:echo_phase}
\begin{figure}[!h]
    \centering
    \includegraphics{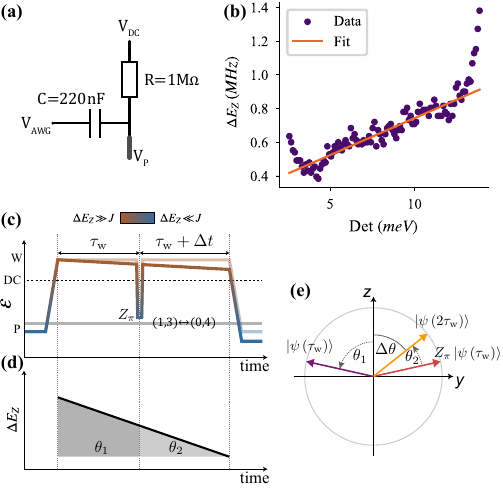}
    \caption{Phase shift analysis in the Hahn-echo experiment. (a) Circuit diagram of the bias-tee used to combine DC bias ($V_{\textnormal{DC}}$) and AC pulses ($V_{\textnormal{AWG}}$) on the plunger gates. (b) Measured dependence of the qubit energy splitting $\Delta E_\textnormal{Z}$ on detuning $\varepsilon$ in the (1,3) charge region (purple dots). The solid orange line indicates a linear fit. Deviations from linearity at the edges arise from increased exchange interaction $J$ near the (0,4) (left) and (2,2) (right) charge transitions.(c) Schematic of the detuning pulse sequence. Due to the finite $RC$ time constant, the actual detuning (solid lines) drifts linearly toward the DC baseline (black dashed line) compared to the ideal square pulse (faint background lines). (d) Time evolution of $\Delta E_\textnormal{Z}$ resulting from the voltage drift. The shaded areas $\theta_1$ and $\theta_2$ represent the accumulated phase during the first and second free-evolution periods, respectively. (e) Qubit state evolution on the $zy$ cross-section of the Bloch sphere. The state vectors correspond to $t=\tau_\textnormal{w}$ (purple), immediately after the $Z_\pi$ refocusing pulse (red), and at the final time $t=2\tau_\textnormal{w}$ (yellow). The non-zero angle $\Delta \theta$ illustrates the phase error caused by the drift.
    }
    \label{fig:sup_fig_echo}
\end{figure}

To apply both DC voltages ($V_{\textnormal{DC}}$) and arbitrary waveform signals ($V_{\textnormal{AWG}}$) to the plunger gates we use a bias-tee, as depicted in Fig.~\ref{fig:sup_fig_echo}(a). The response of the bias-tee is characterized by its $RC$ time constant, $\tau$. In our setup, this time constant is $\tau \approx \SI{220}{\milli\second}$. The bias-tee acts as a high-pass filter. Ideally, high-frequency signals pass through unmodified, satisfying $V_\textnormal{P} \approx V_\textnormal{DC} + V_\textnormal{AWG}$. However, for a step function of amplitude $V_0$ applied at $V_{\textnormal{AWG}}$, the AC component at the plunger gate decays exponentially. Thus, the total voltage evolves as $V_{\textnormal{P}}(t) = V_{\textnormal{DC}} + V_0 \exp(-t / \tau)$.

In our Hahn-echo experiments, we employ pulse sequences with durations on the order of milliseconds. Since these durations are non-negligible compared to $\tau$, the decay of the low-frequency components becomes apparent. Because the detuning $\varepsilon$ is linearly proportional to the gate voltage, this voltage decay translates directly into a time-dependent detuning. Using the approximation $t \ll \tau$, and defining $\varepsilon_0$ as the intended pulse amplitude relative to the DC baseline $\varepsilon_{\textnormal{DC}}$, the detuning follows a linear drift:
\begin{equation}
    \varepsilon(t) \approx \varepsilon_{\textnormal{DC}} + \varepsilon_0 \left( 1 - \frac{t}{\tau} \right).
\end{equation}
This effect is visualized in Fig.~\ref{fig:sup_fig_echo}(c), instead of remaining constant at the operating point W for the sequence duration $2\tau_\textnormal{w}$ (excluding the short refocusing pulse), the detuning drifts linearly back toward the DC baseline.

This detuning drift induces a phase error because the qubit energy splitting, $\Delta E_\textnormal{Z}$, is detuning dependent. We measured $\Delta E_\textnormal{Z}$ across the (1,3) charge region and observed a linear dependence on detuning, as shown in Fig.~\ref{fig:sup_fig_echo}(b). Consequently, the linear drift in detuning maps to a linear decay in the qubit energy splitting over time:
\begin{equation}
    \Delta E_\textnormal{Z}(t) \approx \Delta E_{\textnormal{Z},0} - \alpha t.
\end{equation}
Here, $\alpha= h \beta \varepsilon_0 / \tau$  represents the rate of energy drift, where $\beta$ denotes the linear slope of the qubit frequency dependence on detuning.

\begin{figure*}[t]
    \centering
    \includegraphics{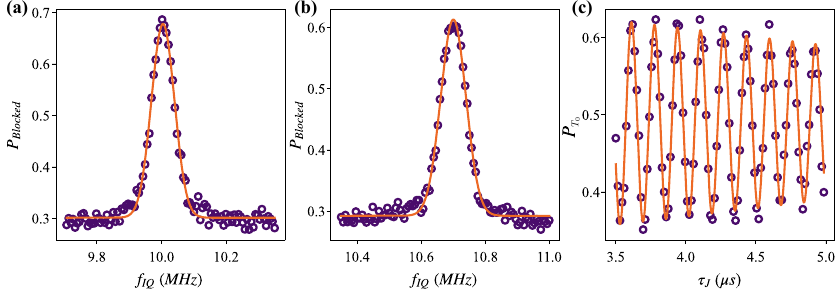}
    \caption{Calibration measurements used to define the Bayesian likelihood models. Purple circles represent data averaged over 3000 single-shot measurements; orange lines indicate the model fits. (a-b) EDSR spectroscopy of spin qubits Q1 and Q2, respectively, plotted against the IQ modulation frequency $f_{\textnormal{IQ}}$. (c) Exchange oscillations as a function of interaction time $\tau_J$.}
    \label{fig:sup_fig_noise}
\end{figure*}

The impact of this energy drift on the qubit dynamics is visualized on the $zy$ cross-section of the Bloch sphere in Fig.~\ref{fig:sup_fig_echo}(e). During the first free-evolution period ($0 \leq t < \tau_\textnormal{w}$), the qubit accumulates a dynamic phase $\theta_1 = \hbar^{-1} \int_0^{\tau_\textnormal{w}} \Delta E_\textnormal{Z}(t) dt$, rotating the state to the purple vector. This evolution is interrupted by a $Z_\pi$ refocusing pulse, which flips the state to the red vector. Finally, during the second period ($\tau_\textnormal{w} \leq t < 2\tau_\textnormal{w}$), the qubit accumulates phase $\theta_2 = \hbar^{-1} \int_{\tau_\textnormal{w}}^{2 \tau_\textnormal{w}} \Delta E_\textnormal{Z}(t) dt$, evolving to the final state shown in yellow.

Ideally, if $\Delta E_\textnormal{Z}$ were constant, $\theta_1$ would equal $\theta_2$, and the final state would align with the z-axis. However, due to the drift, $\theta_1 \neq \theta_2$. The phase error, $\Delta \theta = \theta_1 - \theta_2$, causes the observed phase shift. Substituting the linear approximation for energy drift, we derive the phase error scaling:
\begin{equation}
    \Delta \theta \approx \frac{\alpha \tau_\textnormal{w}^2}{\hbar}.
\end{equation}
We can estimate the evolution time $\tau_\textnormal{w}$ required to induce a phase shift of $\Delta \theta = \pi$. Using the experimental parameters extracted from Fig.~\ref{fig:sup_fig_echo}(b) and the known time constant, we calculate a theoretical value of $\tau_\textnormal{w} \approx \SI{1.76}{\milli\second}$. This is in reasonable agreement with the value of $\SI{3.5}{\milli\second}$ extracted from the Hahn-echo oscillations in Fig.~\ref{fig:fig4}(b), considering the approximations made regarding the linearity of the energy spectrum and nominal tolerances of the $RC$ components.

\section{Noise Characterization and Bayesian Estimation}
\label{app:noise}

We employ a Bayesian estimation framework to simultaneously monitor the temporal fluctuations of three physical quantities: the individual spin qubit frequencies ($f_{\textnormal{Q1}}$, $f_{\textnormal{Q2}}$) and the exchange frequency ($f_\textnormal{J}$). We first describe the specific experiments used to acquire these quantities, followed by the interleaved data acquisition protocol, and finally the Bayesian estimation method details.

\subsection{Experiments and Models}
To track the three variables of interest, we define three distinct experiments. The result of the measuremnt in each experiment is modeled by a probability function $p(x; \theta)$, representing the probability of measuring the blocked state given a sweep parameter $x$ and the physical parameter of interest $\theta$. The constants in these models (visibility, offset, decay times etc.) were determined by averaging results of 3000 single shot measurements [shown in Fig.~\ref{fig:sup_fig_noise}].
 
To measure $f_{\textnormal{Q1}}$ and $f_{\textnormal{Q2}}$, we utilize Electron Dipole Spin Resonance (EDSR). Due to a broken ESR antenna, the spins were driven electrically, limiting the achievable Rabi frequencies to $\SI{50}{\kilo\hertz}$ for Q1 and $\SI{42}{\kilo\hertz}$ for Q2. Consequently, we employed Rabi spectroscopy rather than a Ramsey interferometry measurement. Since the bandwidth of Ramsey fringes is proportional to the Rabi frequency, large fluctuations in the qubit frequency could completely reduce the visibility of Ramsey oscillations. By contrast, Rabi spectroscopy sweeps the drive frequency directly, ensuring that the resonance remains detectable even in the presence of larger qubit frequency fluctuations.

In this method, the duration of the driving pulse is fixed to the calibrated $\pi$-time (\SI{10}{\micro\second} and \SI{12}{\micro\second} respectively). The sweep parameter $x$ is the IQ modulation frequency $f_{\text{IQ}}$. The response is modeled as a Gaussian peak centered at the qubit frequency $\theta = f_0$:
\begin{equation}
    p(f_{\text{IQ}}; f_0) = a + b \exp\left( -\frac{(f_{\text{IQ}} - f_0)^2}{c} \right).
    \label{eqn:model_rabi_spectroscopy}
\end{equation}
The calibration data and fits are displayed in Fig.~\ref{fig:sup_fig_noise}(a) and (b). Note that the total applied frequency is $f_{\text{RF}} = f_{\text{LO}} + f_{\text{IQ}}$.
 
To measure $f_\textnormal{J}$, we perform exchange oscillations measurements described in section~\ref{sec:exchange}. The sweep parameter is the evolution time $x = \tau_J$, and the parameter of interest is the exchange frequency $\theta = f_\textnormal{J}$. The measurement result shows coherent oscillations modeled as:
\begin{equation}
    p(\tau_J; f_J) = a + b \cos(2\pi f_J \tau_J) \exp\left( - (\tau_J / T_2^*)^2 \right).
    \label{eqn:model_exchange_spectroscopy}
\end{equation}
The calibration fit is shown in Fig.~\ref{fig:sup_fig_noise}(c). The constants determined from these fits were held fixed during the Bayesian estimation.

\subsection{Measurement Protocol}
To achieve simultaneous tracking of these parameters, we employed an interleaved measurement scheme~\cite{yoneda2023noise}. Each of the three experiments defined above consists of $N_\textnormal{S} = 100$ distinct sweep points. We group the experimental results into batches where a single batch represents one complete pass through the parameter space for all three experiments, totaling $3 N_\textnormal{S} = 300$ single-shot measurements. To accommodate the requirements of the data acquisition instrument, every single-shot experiment was padded to have an identical duration of \SI{108.8}{\micro\second}, independent of the sweep parameter or experiment type. Therefore, the total acquisition time for a ''batch'' is \SI{32.6}{\milli\second}.

These single-shot measurements are arranged in a strictly alternating sequence, rather than completing a full sweep for one experiment before moving to the next. The acquisition iterates through the sweep index $k$ from $1$ to $N_S$. At each step $k$, the system performs a three single-shot measurements: the RF frequency is set to the $k$-th value for the Q1 sweep, followed immediately by the $k$-th frequency point for Q2, and finally the $k$-th evolution time for the exchange experiment. Once three measurements are done for a given index, the control parameters are updated to $k+1$, and the process repeats until the entire batch is completed. Upon completion, the index resets to $k=1$ to initiate the subsequent batch collection. This protocol was then continuously executed for 5 hours.

\subsection{Bayesian Estimation Framework}
To extract the frequency estimators with high bandwidth, we utilize Bayesian estimation. This allows us to aggregate data over a small number of batches ($R=10$) to obtain a satisfactory uncertainty, resulting in an effective sampling rate of approximately $3\,\text{Hz}$.

Let $j \in \{1, 2, 3\}$ index the experiment (corresponding to $f_{\textnormal{Q1}}$, $f_{\textnormal{Q2}}$, and $f_\textnormal{J}$). The dataset for the $j$-th experiment is denoted:
\begin{equation}
    \mathcal{D}_j = \{(x_{j,k}, m_{j,k}^{(r)}) \mid k=1\dots N_S, \, r=1\dots R\},
\end{equation}
where $x_{j,k}$ is the sweep parameter for the $k$-th step defined in the previous section, and $m_{j,k}^{(r)} \in \{0, 1\}$ is the binary outcome of the single-shot measurement in the $r$-th batch.

Bayes' theorem relates the posterior probability distribution of the parameter $\theta$ to the probability of observing the dataset: $P(\theta | \mathcal{D}) \propto P(\mathcal{D} | \theta) P(\theta)$. We assume a uniform prior distribution $P(\theta)$ over the sweep range. Assuming statistical independence between single-shot measurements, the log-posterior distribution $\mathcal{L}(\theta_j) = \ln P(\theta_j | \mathcal{D}_j)$ is updated as follows:
\begin{equation}
    \mathcal{L}(\theta_j) \propto \sum_{r=1}^{R} \sum_{k=1}^{N_S} \ln P(m_{j,k}^{(r)} | \theta_j, x_{j,k}).
\end{equation}
The term inside the summation is the log-probability of a single-shot experiment. Using the Bernoulli distribution and the physical models $p(x;\theta)$ defined in Eqs.~\ref{eqn:model_rabi_spectroscopy} and ~\ref{eqn:model_exchange_spectroscopy}, this is given by:
\begin{equation}
    \ln P(m | \theta, x) = m \ln[p(x; \theta)] + (1-m) \ln[1 - p(x; \theta)].
\end{equation}
The value of $\theta_j$ that maximizes $\mathcal{L}(\theta_j)$ is selected as the estimator for that parameter step.

\section{SET noise}
\label{app:set_noise}

\begin{figure}[!h]
    \centering
    \includegraphics{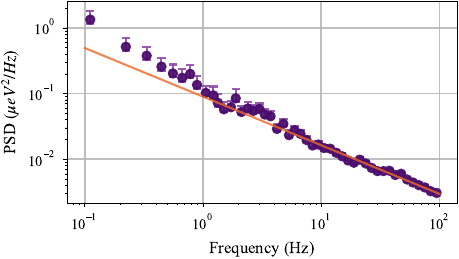}
    \caption{
    Power spectral density (PSD) of the SET used for charge detection [Fig.~\ref{fig:fig1}(a)]. The data (purple circles) represent the intrinsic device noise after subtracting the background IV converter noise floor. The solid orange line indicates a $S_0/f^\alpha$ power-law fit. We estimate a charge noise amplitude $\sqrt{S_0} \approx \SI{0.3}{\micro\electronvolt \per \sqrt{\hertz}}$.
    }
    \label{fig:sup_fig_set_noise}
\end{figure}

We characterize the charge noise spectrum of the left SET [Fig.~\ref{fig:fig1}(a)] by measuring the current fluctuations while biasing the SET at the flank of the Coulomb peak. To isolate the intrinsic device noise, we measure the background noise floor in the Coulomb blockade regime and subtract this contribution from the on-peak signal. The resulting power spectral density (PSD) is plotted in Fig.~\ref{fig:sup_fig_set_noise}. The spectrum deviates from a unique $S_0/f^\alpha$ power law (indicated by the solid orange line). Consequently, rather than relying on the fit extrapolation, we estimate the noise magnitude directly from the measured PSD at \SI{1}{\hertz}. We find a power spectral density of $S(\SI{1}{\hertz}) \approx \SI{0.1}{\micro\electronvolt^2 \per \hertz}$, corresponding to a charge noise amplitude of $\sqrt{S_0} \approx \SI{0.3}{\micro\electronvolt \per \sqrt{\hertz}}$.

\section{Exchange--Qubit Correlations}
\label{app:exchange_correlations}

\begin{figure}[h]
    \centering
    \includegraphics{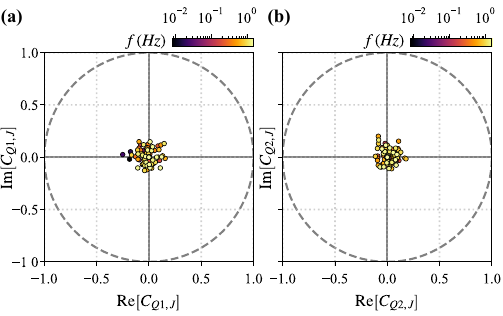}
    \caption{
    Complex correlation coefficient between the exchange frequency $f_\textnormal{J}$ and the individual qubit frequencies. (a) Correlation with $f_{\textnormal{Q1}}$. (b) Correlation with $f_{\textnormal{Q2}}$. The data points cluster around the origin, indicating no significant correlation. The dashed gray circle represents the limit $|C|=1$. The color scale indicates the frequency $f$.
    }
    \label{fig:sup_fig_ex_corr}
\end{figure}

We observe no significant correlations between the fluctuations of the exchange frequency $f_\textnormal{J}$ and the individual qubit frequencies $f_{\textnormal{Q1}}$ and $f_{\textnormal{Q2}}$ within the uncertainty of the Bayesian estimation~\cite{gutierrez2022bayesian}. This is demonstrated in Fig.~\ref{fig:sup_fig_ex_corr}(a) and (b), which display the complex-valued correlation coefficient calculated between the exchange frequency and Q1 ($C_{Q1,J}$), and between exchange and Q2 ($C_{Q2,J}$), respectively. We present the data in the complex plane rather than plotting the magnitude of the correlator because the magnitude is a biased estimator for low correlations. Noise in the real and imaginary components rectifies into a finite positive value, effectively creating a noise floor that mimics weak correlation. The data points for all frequencies (encoded by the color scale) cluster around the origin, indicating the absence of correlation.

\section{Spin-Valley Coupling Model}
\label{app:spin_valley}

We model the dependence of the S--T$_0$ qubit frequency on the external magnetic field $B$ to extract the system parameters. These measurements are performed in the regime of negligible exchange interaction ($J \approx 0$), where the qubit frequency is dominated by the Zeeman energy difference $\Delta E_Z = \Delta g \mu_B B$. Therefore, the S--T$_0$ qubit frequency  follows a linear trend in $B$ determined by the difference in g-factors ($\Delta g$). However, as the magnetic field increases, the Zeeman energy brings the S--T$_0$ qubit states into resonance with states having different valley configurations. This results in avoided crossings mediated by spin-valley coupling ($\Delta_{SV}$), which deviate the qubit frequency from the linear $\Delta E_Z$ trend.

We first analyze the spectrum where both electrons forming the S--T$_0$ qubit occupy the ground valley configuration. The relevant subspace spans the qubit basis $\{ \ket{\uparrow_g \downarrow_g}, \ket{\downarrow_g \uparrow_g} \}$ (where subscripts $g$ and $e$ denote ground and excited valley, respectively) and the $\ket{T_-}$ states involving the excited valley, $\{ \ket{\downarrow_e \downarrow_g}, \ket{\downarrow_g \downarrow_e} \}$. The coupling between these states is described by the Hamiltonian $H_{\textnormal{g}}$:
\begin{equation*}
\resizebox{0.95\linewidth}{!}{$
    H_{\textnormal{g}} = 
    \begin{pmatrix}
    \frac{\Delta g^{gg} \mu_B B}{2} & 0 & \Delta_{SV}^{(1)} & 0 \\
    0 & -\frac{\Delta g^{gg} \mu_B B}{2} & 0 & \Delta_{SV}^{(2)} \\
    \Delta_{SV}^{(1)} & 0 & E_V^{eg} - \bar{g}\mu_B B & 0 \\
    0 & \Delta_{SV}^{(2)} & 0 & E_V^{ge} - \bar{g}\mu_B B
    \end{pmatrix}
$}
\end{equation*}
where $\Delta g^{gg}$ is the g-factor difference in the ground valley configuration, $\bar{g}$ is the average g-factor, $E_V^{eg/ge}$ are the valley splitting energies for the two dots, and $\Delta_{SV}^{(1,2)}$ represents the spin-valley coupling strength in each dot. The solid red line in Fig.~\ref{fig:fig2}(b) is obtained by numerically diagonalizing $H_{\textnormal{g}}$ and plotting the energy difference between the two hybridized S--T$_0$ states ($\ket{\widetilde{\uparrow_g \downarrow_g}}$ and $\ket{\widetilde{\downarrow_g \uparrow_g}}$). The avoided crossings observed at higher fields correspond to the hybridization of the S--T$_0$ states with the $\ket{\downarrow_e \downarrow_g}$ and $\ket{\downarrow_g \downarrow_e}$ states.

To describe the magnetic field dependence related to excited valley S--T$_0$ qubit, we explore the behavior of an S--T$_0$ qubit formed when one electron occupies an excited valley state. We model the interaction of this excited-valley qubit subspace with polarized spin states arising from different valley configurations. The basis states for this configuration are $\{ \ket{\downarrow_e \uparrow_g}, \ket{\uparrow_e \downarrow_g}, \ket{\uparrow_g \uparrow_g}, \ket{\downarrow_e \downarrow_e} \}$. The Hamiltonian $H_{\text{ev}}$ is modeled as:
\begin{equation*}
\resizebox{0.95\linewidth}{!}{$
    H_{\textnormal{e}} = 
    \begin{pmatrix}
    E_V^{eg} + \frac{\Delta g^{eg} \mu_B B}{2} & 0 & 0 & 0 \\
    0 & E_V^{eg} - \frac{\Delta g^{eg} \mu_B B}{2} & \Delta_{SV}^{(1)} & \Delta_{SV}^{(2)} \\
    0 & \Delta_{SV}^{(1)} & \bar{g}\mu_B B & 0 \\
    0 & \Delta_{SV}^{(2)} & 0 & E_V^{ee} - \bar{g}\mu_B B
    \end{pmatrix}
$}
\end{equation*}
Here, $\Delta g^{eg}$ is the g-factor difference for the mixed-valley configuration. The solid blue line in Fig.~\ref{fig:fig2}(b) is obtained by numerically diagonalizing $H_{\textnormal{e}}$ and plotting the energy difference between the $\ket{\downarrow_e \uparrow_g}$ state and hybridized $\ket{\widetilde{\uparrow_e \downarrow_g}}$ state.

\bibliography{references.bib}

@article{loss1998quantum,
  title={Quantum computation with quantum dots},
  author={Loss, Daniel and DiVincenzo, David P},
  journal={Physical Review A},
  volume={57},
  number={1},
  pages={120},
  year={1998},
  publisher={APS}
}

@article{nonergodic_noise,
  title={Quantum dephasing in a gated GaAs triple quantum dot due to nonergodic noise},
  author={Delbecq, MR and Nakajima, T and Stano, P and Otsuka, T and Amaha, S and Yoneda, J and Takeda, K and Allison, G and Ludwig, A and Wieck, AD and others},
  journal={Physical review letters},
  volume={116},
  number={4},
  pages={046802},
  year={2016},
  publisher={APS}
}

@article{yoneda2023noise,
  title={Noise-correlation spectrum for a pair of spin qubits in silicon},
  author={Yoneda, J and Rojas-Arias, JS and Stano, P and Takeda, K and Noiri, A and Nakajima, T and Loss, D and Tarucha, S},
  journal={Nature Physics},
  volume={19},
  number={12},
  pages={1793--1798},
  year={2023},
  publisher={Nature Publishing Group UK London}
}

@article{gutierrez2022bayesian,
  title={Bayesian estimation of correlation functions},
  author={Guti{\'e}rrez-Rubio, {\'A}ngel and Rojas-Arias, Juan S and Yoneda, Jun and Tarucha, Seigo and Loss, Daniel and Stano, Peter},
  journal={Physical Review Research},
  volume={4},
  number={4},
  pages={043166},
  year={2022},
  publisher={APS}
}

@article{lisenfeld2023enhancing,
  title={Enhancing the coherence of superconducting quantum bits with electric fields},
  author={Lisenfeld, J{\"u}rgen and Bilmes, Alexander and Ustinov, Alexey V},
  journal={npj Quantum Information},
  volume={9},
  number={1},
  pages={8},
  year={2023},
  publisher={Nature Publishing Group UK London}
}

@article{cifuentes2024bounds,
  title={Bounds to electron spin qubit variability for scalable CMOS architectures},
  author={Cifuentes, Jes{\'u}s D and Tanttu, Tuomo and Gilbert, Will and Huang, Jonathan Y and Vahapoglu, Ensar and Leon, Ross CC and Serrano, Santiago and Otter, Dennis and Dunmore, Daniel and Mai, Philip Y and others},
  journal={Nature communications},
  volume={15},
  number={1},
  pages={4299},
  year={2024},
  publisher={Nature Publishing Group UK London}
}

@article{jock2018silicon,
  title={A silicon metal-oxide-semiconductor electron spin-orbit qubit},
  author={Jock, Ryan M and Jacobson, N Tobias and Harvey-Collard, Patrick and Mounce, Andrew M and Srinivasa, Vanita and Ward, Dan R and Anderson, John and Manginell, Ron and Wendt, Joel R and Rudolph, Martin and others},
  journal={Nature communications},
  volume={9},
  number={1},
  pages={1768},
  year={2018},
  publisher={Nature Publishing Group UK London}
}

@article{friesen2007valley,
  title={Valley splitting theory of Si Ge/ Si/ Si Ge quantum wells},
  author={Friesen, Mark and Chutia, Sucismita and Tahan, Charles and Coppersmith, SN},
  journal={Physical Review B—Condensed Matter and Materials Physics},
  volume={75},
  number={11},
  pages={115318},
  year={2007},
  publisher={APS}
}

@article{gonzalez2021scaling,
  title={Scaling silicon-based quantum computing using CMOS technology},
  author={Gonzalez-Zalba, MF and De Franceschi, S and Charbon, E and Meunier, Tristan and Vinet, M and Dzurak, AS},
  journal={Nature Electronics},
  volume={4},
  number={12},
  pages={872--884},
  year={2021},
  publisher={Nature Publishing Group UK London}
}

@article{zwerver2022qubits,
  title={Qubits made by advanced semiconductor manufacturing},
  author={Zwerver, AMJ and Kr{\"a}henmann, T and Watson, TF and Lampert, Lester and George, Hubert C and Pillarisetty, Ravi and Bojarski, SA and Amin, Payam and Amitonov, SV and Boter, JM and others},
  journal={Nature Electronics},
  volume={5},
  number={3},
  pages={184--190},
  year={2022},
  publisher={Nature Publishing Group UK London}
}

@article{kane1998silicon,
  title={A silicon-based nuclear spin quantum computer},
  author={Kane, Bruce E},
  journal={nature},
  volume={393},
  number={6681},
  pages={133--137},
  year={1998},
  publisher={Nature Publishing Group UK London}
}

@article{zwanenburg2013silicon,
  title={Silicon quantum electronics},
  author={Zwanenburg, Floris A and Dzurak, Andrew S and Morello, Andrea and Simmons, Michelle Y and Hollenberg, Lloyd CL and Klimeck, Gerhard and Rogge, Sven and Coppersmith, Susan N and Eriksson, Mark A},
  journal={Reviews of modern physics},
  volume={85},
  number={3},
  pages={961--1019},
  year={2013},
  publisher={APS}
}

@article{vandersypen2019quantum,
  title={Quantum computing with semiconductor spins},
  author={Vandersypen, Lieven MK and Eriksson, Mark A},
  journal={Physics Today},
  volume={72},
  number={8},
  pages={38--45},
  year={2019},
  publisher={AIP Publishing}
}

@article{maurand2016cmos,
  title={A CMOS silicon spin qubit},
  author={Maurand, R and Jehl, X and Kotekar-Patil, D and Corna, Andrea and Bohuslavskyi, Heorhii and Lavi{\'e}ville, R and Hutin, L and Barraud, S and Vinet, M and Sanquer, M and others},
  journal={Nature communications},
  volume={7},
  number={1},
  pages={13575},
  year={2016},
  publisher={Nature Publishing Group UK London}
}

@article{veldhorst2014addressable,
  title={An addressable quantum dot qubit with fault-tolerant control-fidelity},
  author={Veldhorst, M and Hwang, JCC and Yang, CH and Leenstra, AW and de Ronde, Bob and Dehollain, JP and Muhonen, JT and Hudson, FE and Itoh, Kohei M and Morello, A t and others},
  journal={Nature nanotechnology},
  volume={9},
  number={12},
  pages={981--985},
  year={2014},
  publisher={Nature Publishing Group UK London}
}

@article{xue2022quantum,
  title={Quantum logic with spin qubits crossing the surface code threshold},
  author={Xue, Xiao and Russ, Maximilian and Samkharadze, Nodar and Undseth, Brennan and Sammak, Amir and Scappucci, Giordano and Vandersypen, Lieven MK},
  journal={Nature},
  volume={601},
  number={7893},
  pages={343--347},
  year={2022},
  publisher={Nature Publishing Group UK London}
}

@article{noiri2022fast,
  title={Fast universal quantum gate above the fault-tolerance threshold in silicon},
  author={Noiri, Akito and Takeda, Kenta and Nakajima, Takashi and Kobayashi, Takashi and Sammak, Amir and Scappucci, Giordano and Tarucha, Seigo},
  journal={Nature},
  volume={601},
  number={7893},
  pages={338--342},
  year={2022},
  publisher={Nature Publishing Group UK London}
}

@article{mkadzik2022precision,
  title={Precision tomography of a three-qubit donor quantum processor in silicon},
  author={M{\k{a}}dzik, Mateusz T and Asaad, Serwan and Youssry, Akram and Joecker, Benjamin and Rudinger, Kenneth M and Nielsen, Erik and Young, Kevin C and Proctor, Timothy J and Baczewski, Andrew D and Laucht, Arne and others},
  journal={Nature},
  volume={601},
  number={7893},
  pages={348--353},
  year={2022},
  publisher={Nature Publishing Group UK London}
}

@article{harvey2022coherent,
  title={Coherent spin-spin coupling mediated by virtual microwave photons},
  author={Harvey-Collard, Patrick and Dijkema, Jurgen and Zheng, Guoji and Sammak, Amir and Scappucci, Giordano and Vandersypen, Lieven MK},
  journal={Physical Review X},
  volume={12},
  number={2},
  pages={021026},
  year={2022},
  publisher={APS}
}

@article{de2025high,
  title={High-fidelity single-spin shuttling in silicon},
  author={De Smet, Maxim and Matsumoto, Yuta and Zwerver, Anne-Marije J and Tryputen, Larysa and de Snoo, Sander L and Amitonov, Sergey V and Katiraee-Far, Sam R and Sammak, Amir and Samkharadze, Nodar and G{\"u}l, {\"O}nder and others},
  journal={Nature Nanotechnology},
  pages={1--7},
  year={2025},
  publisher={Nature Publishing Group UK London}
}

@article{steinacker2025industry,
  title={Industry-compatible silicon spin-qubit unit cells exceeding 99\% fidelity},
  author={Steinacker, Paul and Dumoulin Stuyck, Nard and Lim, Wee Han and Tanttu, Tuomo and Feng, MengKe and Serrano, Santiago and Nickl, Andreas and Candido, Marco and Cifuentes, Jesus D and Vahapoglu, Ensar and others},
  journal={Nature},
  pages={1--7},
  year={2025},
  publisher={Nature Publishing Group UK London}
}

@article{petit2020universal,
  title={Universal quantum logic in hot silicon qubits},
  author={Petit, Luca and Eenink, HGJ and Russ, M and Lawrie, WIL and Hendrickx, NW and Philips, SGJ and Clarke, JS and Vandersypen, LMK and Veldhorst, M},
  journal={Nature},
  volume={580},
  number={7803},
  pages={355--359},
  year={2020},
  publisher={Nature Publishing Group UK London}
}

@article{stanoloss2025,
  title={Review of performance metrics of spin qubits in gated semiconducting nanostructures},
  author={Stano, Peter and Loss, Daniel},
  journal={arXiv preprint arXiv:2107.06485v9},
  year={2025}
}

@article{petta2005coherent,
  title={Coherent manipulation of coupled electron spins in semiconductor quantum dots},
  author={Petta, Jason R and Johnson, Alexander Comstock and Taylor, Jacob M and Laird, Edward A and Yacoby, Amir and Lukin, Mikhail D and Marcus, Charles M and Hanson, Micah P and Gossard, Arthur C},
  journal={Science},
  volume={309},
  number={5744},
  pages={2180--2184},
  year={2005},
  publisher={American Association for the Advancement of Science}
}

@article{dial2013charge,
  title={Charge noise spectroscopy using coherent exchange oscillations in a singlet-triplet qubit},
  author={Dial, OE and Shulman, Michael Dean and Harvey, Shannon Pasca and Bluhm, H and Umansky, V and Yacoby, Amnon},
  journal={Physical review letters},
  volume={110},
  number={14},
  pages={146804},
  year={2013},
  publisher={APS}
}

@article{cywinski2008enhance,
  title={How to enhance dephasing time in superconducting qubits},
  author={Cywi{\'n}ski, {\L}ukasz and Lutchyn, Roman M and Nave, Cody P and Das Sarma, S},
  journal={Physical Review B—Condensed Matter and Materials Physics},
  volume={77},
  number={17},
  pages={174509},
  year={2008},
  publisher={APS}
}

@article{yoneda2018quantum,
  title={A quantum-dot spin qubit with coherence limited by charge noise and fidelity higher than 99.9\%},
  author={Yoneda, Jun and Takeda, Kenta and Otsuka, Tomohiro and Nakajima, Takashi and Delbecq, Matthieu R and Allison, Giles and Honda, Takumu and Kodera, Tetsuo and Oda, Shunri and Hoshi, Yusuke and others},
  journal={Nature nanotechnology},
  volume={13},
  number={2},
  pages={102--106},
  year={2018},
  publisher={Nature Publishing Group UK London}
}

@article{wu2025simultaneous,
  title={Simultaneous high-fidelity single-qubit gates in a spin qubit array},
  author={Wu, Yi-Hsien and Camenzind, Leon C and B{\"u}tler, Patrick and Jin, Ik Kyeong and Noiri, Akito and Takeda, Kenta and Nakajima, Takashi and Kobayashi, Takashi and Scappucci, Giordano and Goan, Hsi-Sheng and others},
  journal={arXiv preprint arXiv:2507.11918},
  year={2025}
}

@article{yang2019silicon,
  title={Silicon qubit fidelities approaching incoherent noise limits via pulse engineering},
  author={Yang, CH and Chan, KW and Harper, Robin and Huang, Wister and Evans, T and Hwang, JCC and Hensen, B and Laucht, Arne and Tanttu, Tuomo and Hudson, FE and others},
  journal={Nature Electronics},
  volume={2},
  number={4},
  pages={151--158},
  year={2019},
  publisher={Nature Publishing Group UK London}
}

@article{mills2022two,
  title={Two-qubit silicon quantum processor with operation fidelity exceeding 99\%},
  author={Mills, Adam R and Guinn, Charles R and Gullans, Michael J and Sigillito, Anthony J and Feldman, Mayer M and Nielsen, Erik and Petta, Jason R},
  journal={Science Advances},
  volume={8},
  number={14},
  pages={eabn5130},
  year={2022},
  publisher={American Association for the Advancement of Science}
}

@article{struck2024spin,
  title={Spin-EPR-pair separation by conveyor-mode single electron shuttling in Si/SiGe},
  author={Struck, Tom and Volmer, Mats and Visser, Lino and Offermann, Tobias and Xue, Ran and Tu, Jhih-Sian and Trellenkamp, Stefan and Cywi{\'n}ski, {\L}ukasz and Bluhm, Hendrik and Schreiber, Lars R},
  journal={Nature Communications},
  volume={15},
  number={1},
  pages={1325},
  year={2024},
  publisher={Nature Publishing Group UK London}
}

@article{xue2024si,
  title={Si/SiGe QuBus for single electron information-processing devices with memory and micron-scale connectivity function},
  author={Xue, Ran and Beer, Max and Seidler, Inga and Humpohl, Simon and Tu, Jhih-Sian and Trellenkamp, Stefan and Struck, Tom and Bluhm, Hendrik and Schreiber, Lars R},
  journal={Nature Communications},
  volume={15},
  number={1},
  pages={2296},
  year={2024},
  publisher={Nature Publishing Group UK London}
}

@article{harvey2018high,
  title={High-fidelity single-shot readout for a spin qubit via an enhanced latching mechanism},
  author={Harvey-Collard, Patrick and D’Anjou, Benjamin and Rudolph, Martin and Jacobson, N Tobias and Dominguez, Jason and Ten Eyck, Gregory A and Wendt, Joel R and Pluym, Tammy and Lilly, Michael P and Coish, William A and others},
  journal={Physical Review X},
  volume={8},
  number={2},
  pages={021046},
  year={2018},
  publisher={APS}
}

@article{levy2002universal,
  title={Universal quantum computation with spin-1/2 pairs and Heisenberg exchange},
  author={Levy, Jeremy},
  journal={Physical Review Letters},
  volume={89},
  number={14},
  pages={147902},
  year={2002},
  publisher={APS}
}

@article{taylor2007relaxation,
  title={Relaxation, dephasing, and quantum control of electron spins in double quantum dots},
  author={Taylor, JM and Petta, JR and Johnson, AC and Yacoby, Amnon and Marcus, CM and Lukin, MD},
  journal={Physical Review B—Condensed Matter and Materials Physics},
  volume={76},
  number={3},
  pages={035315},
  year={2007},
  publisher={APS}
}

@article{jock2022silicon,
  title={A silicon singlet--triplet qubit driven by spin-valley coupling},
  author={Jock, Ryan M and Jacobson, N Tobias and Rudolph, Martin and Ward, Daniel R and Carroll, Malcolm S and Luhman, Dwight R},
  journal={Nature communications},
  volume={13},
  number={1},
  pages={641},
  year={2022},
  publisher={Nature Publishing Group UK London}
}

@article{tanttu2019controlling,
  title={Controlling spin-orbit interactions in silicon quantum dots using magnetic field direction},
  author={Tanttu, Tuomo and Hensen, Bas and Chan, Kok Wai and Yang, Chih Hwan and Huang, Wister Wei and Fogarty, Michael and Hudson, Fay and Itoh, Kohei and Culcer, Dimitrie and Laucht, Arne and others},
  journal={Physical Review X},
  volume={9},
  number={2},
  pages={021028},
  year={2019},
  publisher={APS}
}

@article{chittock2025radio,
  title={Radio-frequency cascade readout of coupled spin qubits fabricated using a 300~mm wafer process},
  author={Chittock-Wood, Jacob F and Leon, Ross CC and Fogarty, Michael A and Murphy, Tara and Patom{\"a}ki, Sofia M and Oakes, Giovanni A and Williams, James and von Horstig, Felix-Ekkehard and Johnson, Nathan and Jussot, Julien and others},
  journal={arXiv preprint arXiv:2408.01241},
  year={2025}
}

@article{elsayed2024low,
  title={Low charge noise quantum dots with industrial CMOS manufacturing},
  author={Elsayed, Asser and Shehata, MMK and Godfrin, Clement and Kubicek, Stefan and Massar, Shana and Canvel, Yann and Jussot, Julien and Simion, George and Mongillo, Massimo and Wan, Danny and others},
  journal={npj Quantum Information},
  volume={10},
  number={1},
  pages={70},
  year={2024},
  publisher={Nature Publishing Group UK London}
}

@article{petersson2010quantum,
  title={Quantum coherence in a one-electron semiconductor charge qubit},
  author={Petersson, KD and Petta, JR and Lu, H and Gossard, AC},
  journal={Physical Review Letters},
  volume={105},
  number={24},
  pages={246804},
  year={2010},
  publisher={APS}
}

@article{thorgrimsson2017extending,
  title={Extending the coherence of a quantum dot hybrid qubit},
  author={Thorgrimsson, Brandur and Kim, Dohun and Yang, Yuan-Chi and Smith, LW and Simmons, CB and Ward, Daniel R and Foote, Ryan H and Corrigan, J and Savage, DE and Lagally, MG and others},
  journal={npj Quantum Information},
  volume={3},
  number={1},
  pages={32},
  year={2017},
  publisher={Nature Publishing Group UK London}
}

@article{kranz2020exploiting,
  title={Exploiting a single-crystal environment to minimize the charge noise on qubits in silicon},
  author={Kranz, Ludwik and Gorman, Samuel Keith and Thorgrimsson, Brandur and He, Yu and Keith, Daniel and Keizer, Joris Gerhard and Simmons, Michelle Yvonne},
  journal={Advanced Materials},
  volume={32},
  number={40},
  pages={2003361},
  year={2020},
  publisher={Wiley Online Library}
}

@article{jirovec2021singlet,
  title={A singlet-triplet hole spin qubit in planar Ge},
  author={Jirovec, Daniel and Hofmann, Andrea and Ballabio, Andrea and Mutter, Philipp M and Tavani, Giulio and Botifoll, Marc and Crippa, Alessandro and Kukucka, Josip and Sagi, Oliver and Martins, Frederico and others},
  journal={Nature Materials},
  volume={20},
  number={8},
  pages={1106--1112},
  year={2021},
  publisher={Nature Publishing Group UK London}
}

@article{reed2016reduced,
  title={Reduced sensitivity to charge noise in semiconductor spin qubits via symmetric operation},
  author={Reed, MD and Maune, BM and Andrews, RW and Borselli, MG and Eng, K and Jura, MP and Kiselev, AA and Ladd, TD and Merkel, ST and Milosavljevic, I and others},
  journal={Physical review letters},
  volume={116},
  number={11},
  pages={110402},
  year={2016},
  publisher={APS}
}

@article{martins2016noise,
  title={Noise suppression using symmetric exchange gates in spin qubits},
  author={Martins, Frederico and Malinowski, Filip K and Nissen, Peter D and Barnes, Edwin and Fallahi, Saeed and Gardner, Geoffrey C and Manfra, Michael J and Marcus, Charles M and Kuemmeth, Ferdinand},
  journal={Physical review letters},
  volume={116},
  number={11},
  pages={116801},
  year={2016},
  publisher={APS}
}

@article{ha2025two,
  title={Two-dimensional Si spin qubit arrays with multilevel interconnects},
  author={Ha, Sieu D and Acuna, Edwin and Raach, Kate and Bloom, Zachery T and Brecht, Teresa L and Chappell, James M and Choi, Maxwell D and Christensen, Justin E and Counts, Ian T and Daprano, Dominic and others},
  journal={PRX Quantum},
  volume={6},
  number={3},
  pages={030327},
  year={2025},
  publisher={APS}
}

@article{weinstein2023universal,
  title={Universal logic with encoded spin qubits in silicon},
  author={Weinstein, Aaron J and Reed, Matthew D and Jones, Aaron M and Andrews, Reed W and Barnes, David and Blumoff, Jacob Z and Euliss, Larken E and Eng, Kevin and Fong, Bryan H and Ha, Sieu D and others},
  journal={Nature},
  volume={615},
  number={7954},
  pages={817--822},
  year={2023},
  publisher={Nature Publishing Group UK London}
}

@article{paladino20141,
  title={1/f noise: Implications for solid-state quantum information},
  author={Paladino, E and Galperin, YM and Falci, G and Altshuler, BL},
  journal={Reviews of Modern Physics},
  volume={86},
  number={2},
  pages={361--418},
  year={2014},
  publisher={APS}
}

@article{xue2021cmos,
  title={CMOS-based cryogenic control of silicon quantum circuits},
  author={Xue, Xiao and Patra, Bishnu and van Dijk, Jeroen PG and Samkharadze, Nodar and Subramanian, Sushil and Corna, Andrea and Paquelet Wuetz, Brian and Jeon, Charles and Sheikh, Farhana and Juarez-Hernandez, Esdras and others},
  journal={Nature},
  volume={593},
  number={7858},
  pages={205--210},
  year={2021},
  publisher={Nature Publishing Group UK London}
}

@article{bartee2025spin,
  title={Spin-qubit control with a milli-kelvin CMOS chip},
  author={Bartee, Samuel K and Gilbert, Will and Zuo, Kun and Das, Kushal and Tanttu, Tuomo and Yang, Chih Hwan and Dumoulin Stuyck, Nard and Pauka, Sebastian J and Su, Rocky Y and Lim, Wee Han and others},
  journal={Nature},
  pages={1--6},
  year={2025},
  publisher={Nature Publishing Group UK London}
}

@article{fowler2012surface,
  title={Surface codes: Towards practical large-scale quantum computation},
  author={Fowler, Austin G and Mariantoni, Matteo and Martinis, John M and Cleland, Andrew N},
  journal={Physical Review A—Atomic, Molecular, and Optical Physics},
  volume={86},
  number={3},
  pages={032324},
  year={2012},
  publisher={APS}
}

@article{bombin2006topological,
  title={Topological quantum distillation},
  author={Bombin, Hector and Martin-Delgado, Miguel Angel},
  journal={Physical review letters},
  volume={97},
  number={18},
  pages={180501},
  year={2006},
  publisher={APS}
}

@article{mcewen2022resolving,
  title={Resolving catastrophic error bursts from cosmic rays in large arrays of superconducting qubits},
  author={McEwen, Matt and Faoro, Lara and Arya, Kunal and Dunsworth, Andrew and Huang, Trent and Kim, Seon and Burkett, Brian and Fowler, Austin and Arute, Frank and Bardin, Joseph C and others},
  journal={Nature Physics},
  volume={18},
  number={1},
  pages={107--111},
  year={2022},
  publisher={Nature Publishing Group UK London}
}

@article{miao2023overcoming,
  title={Overcoming leakage in quantum error correction},
  author={Miao, Kevin C and McEwen, Matt and Atalaya, Juan and Kafri, Dvir and Pryadko, Leonid P and Bengtsson, Andreas and Opremcak, Alex and Satzinger, Kevin J and Chen, Zijun and Klimov, Paul V and others},
  journal={Nature Physics},
  volume={19},
  number={12},
  pages={1780--1786},
  year={2023},
  publisher={Nature Publishing Group UK London}
}

@article{google2025quantum,
  title={Quantum error correction below the surface code threshold},
  journal={Nature},
  volume={638},
  number={8052},
  pages={920--926},
  year={2025},
  publisher={Nature Publishing Group UK London}
}

@article{veldhorst2015spin,
  title={Spin-orbit coupling and operation of multivalley spin qubits},
  author={Veldhorst, M and Ruskov, R and Yang, CH and Hwang, JCC and Hudson, FE and Flatt{\'e}, ME and Tahan, C and Itoh, Kohei M and Morello, A and Dzurak, AS},
  journal={Physical Review B},
  volume={92},
  number={20},
  pages={201401},
  year={2015},
  publisher={APS}
}

@article{ruskov2018electron,
  title={Electron g-factor of valley states in realistic silicon quantum dots},
  author={Ruskov, Rusko and Veldhorst, Menno and Dzurak, Andrew S and Tahan, Charles},
  journal={Physical Review B},
  volume={98},
  number={24},
  pages={245424},
  year={2018},
  publisher={APS}
}

@article{burkard2023semiconductor,
  title={Semiconductor spin qubits},
  author={Burkard, Guido and Ladd, Thaddeus D and Pan, Andrew and Nichol, John M and Petta, Jason R},
  journal={Reviews of Modern Physics},
  volume={95},
  number={2},
  pages={025003},
  year={2023},
  publisher={APS}
}

\end{document}